\documentclass[fleqn,usenatbib]{mnras}

\usepackage{newtxtext,newtxmath}
\usepackage{lipsum}
\usepackage[T1]{fontenc}
\usepackage{longtable}
\DeclareRobustCommand{\VAN}[3]{#2}
\let\VANthebibliography\thebibliography
\def\thebibliography{\DeclareRobustCommand{\VAN}[3]{##3}\VANthebibliography}

\usepackage{graphicx}	
\usepackage{amsmath}	
\usepackage{pdflscape}

\usepackage{supertabular}
\usepackage{microtype} 
\usepackage{enumitem}

\usepackage{etoolbox, siunitx}
\robustify\bfseries

\usepackage{booktabs} 

\usepackage[normalem]{ulem} 

\DeclareSIUnit\au{AU}
\DeclareSIUnit\Rsun{R_\odot}
\DeclareSIUnit\Rjup{R_\text{Jup}}
\DeclareSIUnit\Msun{M_\odot}
\DeclareSIUnit\Mjup{M_\text{Jup}}
\DeclareSIUnit\gyr{Gyr}
\DeclareSIUnit\ppt{ppt}
\DeclareSIUnit\ppm{ppm}

\title[EBLM active longitudes]{Tight stellar binaries favour active longitudes at sub- and anti-stellar points}

\author[Sethi \& Martin]{%
        Ritika Sethi$^{1,2*}$$^{\href{https://orcid.org/0000-0002-6576-3346}{\includegraphics[scale=0.5]{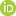}}}$ \& David V. Martin$^{3,4,5}$$^{\href{https://orcid.org/0000-0002-7595-6360}{\includegraphics[scale=0.5]{orcid.jpg}}}$,
\\
$^{1}$MIT Kavli Institute for Astrophysics and Space Research, Massachusetts Institute of Technology, Cambridge, MA 02139, USA\\
$^{2}$Department of Physical Sciences, Indian Institute of Science Education and Research Berhampur, Berhampur 760010, Odisha, India \\
$^{3}$Department of Physics \& Astronomy, Tufts University, Medford, MA 02155, USA \\
$^{4}$Department of Astronomy, The Ohio State University, 4055 McPherson Laboratory, Columbus, OH 43210, USA\\
$^{5}$NASA Sagan Fellow\\
*rsethi@mit.edu\\
}

\date{First submitted to MNRAS Sept 15 2023, Second submission Jan 24 2024, Accepted March 04 2024}

\pubyear{2024}

\begin{document}

\label{firstpage}
\maketitle

\begin{abstract}

Stellar binaries are ubiquitous in the galaxy and a laboratory for astrophysical effects. We use TESS to study photometric modulations in the lightcurves of 162 unequal mass eclipsing binaries from the EBLM (Eclipsing Binary Low Mass) survey, comprising F/G/K primaries and M-dwarf secondaries. We detect modulations on 81 eclipsing binaries. We catalog the rotation rates of the primary star in 69 binaries and discover 17 ellipsoidal variables. In a large portion (at least $\sim 51\%$) of our sample, we detect photometric modulations consistent with two over-densities of spots on the primary star that are roughly $180^{\circ}$ apart. We show that these so-called active longitudes are preferentially at the sub- and anti-stellar points on the primary star. Physically, this means that the spots on the primary star preferentially face directly towards and away from the secondary star.

\end{abstract}




\section{INTRODUCTION}\label{sec:introduction}

Precise measurements of stellar rotation periods give insights into stellar structure and evolution \citep{1953MNRAS.113..716M}. The field of gyrochonology connects a star's age to its rotation rate \citep{Barnes2003,Mamajek2008,Angus2015}. The rotation of a star is also closely linked to its magnetic field generation and activity levels, which causes phenomena such as coronal mass ejections, flares, and  spots. Stellar rotation rates can be determined from spectroscopic line broadening, but they are typically degenerate with the  inclination of the star along the line of sight. With advancements in high-precision, long- timespan photometry in space (e.g. Kepler, TESS), it can be more effective to measure photometric modulations due to the presence of spots on the stellar photosphere. Over the past decade, various methods like Lomb-Scargle periodograms \citep{Lurie2017, 2015A&A...583A..65R, 2019MNRAS.487..304D, 2013A&A...557L..10N}, autocorrelation functions (ACFs) \citep{2014ApJS..211...24M, 2013MNRAS.432.1203M}, wavelet analysis \citep{refId0, 2010A&A...511A..46M, 2014A&A...572A..34G}, and Gaussian processes \citep{2018MNRAS.474.2094A} have been used to identify  rotational signatures. 

There is significant interest in measuring rotation rates for not only single stars but binaries as well. Rotation rates in binaries can be indicative of tidal evolution and synchronization \citep{1981A&A....99..126H, 2003A&A...405..303H,Stassun1999}. This is particularly powerful if combined with other indicators of tidal evolution like orbital circularisation \citep{Meibom2006} and spin-orbit alignment \citep{Albrecht2011,Hodzic2020}. Stellar binarity may feature magnetic interactions that affect rotation and stellar activity \citep{Morales2010,QingFeng2019}. It was proposed by \citet{1988A&A...205..167V, 2000IAUS..200P.217H}  that there could be magnetic filaments interacting between the two stars, which could result in increased flare rates and “active longitudes.” The latter is a phenomenon where specific longitudes of the stellar surface contain more spots than others. \citet{Berdyugina1998} observed active longitudes in four RS CVn binaries. They also observed the ``flip-flop'' effect of the longitude changing every few years \citet{Jetsu1993,Jetsu1994}. \citet{Olah2006} studied active longitudes in 12 eclipsing binaries, three of which have active longitudes at the sub- and anti-stellar points (i.e. an abundance of spots facing towards and away from the other star). \citet{Martin2024} showed likely active longitudes in the eclipsing binary CM Draconis, including evolution over a timescale of years.


The TESS satellite's nearly all-sky coverage has revolutionalized rotation rate measurements \citep{Medina2020,Howard2020,Gilbert2022}. 
However, TESS systematics have made it challenging to find spot modulations longer than the TESS spacecraft's orbital period of 13.7 days owing to confusion between instrumental and astrophysical variability \citep{2022ApJ...930....7A, 2020ApJS..250...20C}. This has been partially mitigated through methods such as deep learning \citep{Claytor2022} and background star analysis \citep{Hedges2020,Martin2021}, but in general it remains a challenge. Another limitation is differential rotation: ignorance of the spot's latitude means we may not be observing the star's equatorial latitude \citep{Lurie2017, 2015A&A...583A..65R}. The measured rotation rate may also change over time as spots evolve. Furthermore, periodic signals from tidal distortion effects like ellipsoidal variations are often confused for rotation signatures. Finally, a challenge of measuring rotation rates in binaries is that if both stars are of similar brightness there may be two spot modulation signals, and we are uncertain to which star they should be ascribed \citep{Martin2024}.

In the present study, we study a sample of 162 F/G/K + M eclipsing binaries from the Eclipsing Binary Low-Mass (EBLM) catalog (henceforth referred to as EBLMs) \citep{Triaud2013,Triaud2017,Gill2019,Swayne2021,Duck2023,Sebastian2023}. The unequal mass ratio means that the difference in visual magnitude is between 4 and 12, with a peak at 8 \citep{Triaud2017}. This means that even in the case of least discrepant flux, the primary star is at least 40 times brighter than the secondary. Barring the unexpected case of an M-dwarf being so spotty that its flux varies by 10's of per cent, only spots on the primary star will be visible. This means we are oblivious to the M-dwarf's rotation rate, but any ambiguity of the spot host is removed. In our sample 90\% of the binaries have a orbital period of less than 14 days. If tidally locked, this means the rotation rates will be amenable to TESS detection. We further preprocess the TESS light curves before rotation period measurement to remove any long-term systematics while retaining the most real variability. The periodic variability in the light curves is detected using three methods: Lomb Scargle periodograms and autocorrelation functions \citep{2013MNRAS.432.1203M, 2014ApJS..211...24M}, and a two-sinusoid method introduced in \citep{Martin2024} in the study of CM Draconis. 

Our paper serves two purposes. First, it is a catalog of rotation rates for the EBLM sample; out of 162 eclipsing binaries, we measure rotation rates in 69. We also identify 17 ellipsoidal variables. This catalog may be used for future studies, such as tidal evolution in unequal mass ratio binaries. Measured rotation rates of the primary may also factor into future calculations of fundamental stellar parameters using eclipsing binaries, which is one of the primary goals of the EBLM project. The second purpose, which we explore in this paper, is the detection of active longitudes in tight binaries. They are measured relative to the orbital phase, indicating the orientation of spots on the primary star relative to the location of the secondary star. We detect active longitudes in 41 of our stars and demonstrate that they prefer to occur typically at the sub- and anti-stellar points.

This paper is organised as follows. First, we describe how we acquire and process the data in Sect.~\ref{sec:data_acquisition}. In Sect.~\ref{sec:methodology}, we discuss the types of signals we are looking for and how we classify them. We then present and discuss our results in Sect.~\ref{sec:results}, before concluding in Sect.~\ref{sec:conclusion}.

\section{Data Acquisition and Preparation}\label{sec:data_acquisition}

The TESS spacecraft, which is now operating on an extended mission, observed a major part of the sky split into 26 sectors, where each sector was observed for two 13.7 day long TESS orbit. The gap in data at the end of each orbit is due to data download to the Earth. We began with 208 EBLM targets and downloaded the PDCSAP TESS light curves corrected for instrumental errors using the SPOC pipeline \citep{2016SPIE.9913E..3EJ} for 162 of them, depending on their availability at the time of conducting this analysis (June 2022). The data download was done using the \textsc{lightkurve} package \citep{2018ascl.soft12013L}. 

\begin{figure}
    \centering
    \includegraphics[width = 0.45\textwidth]{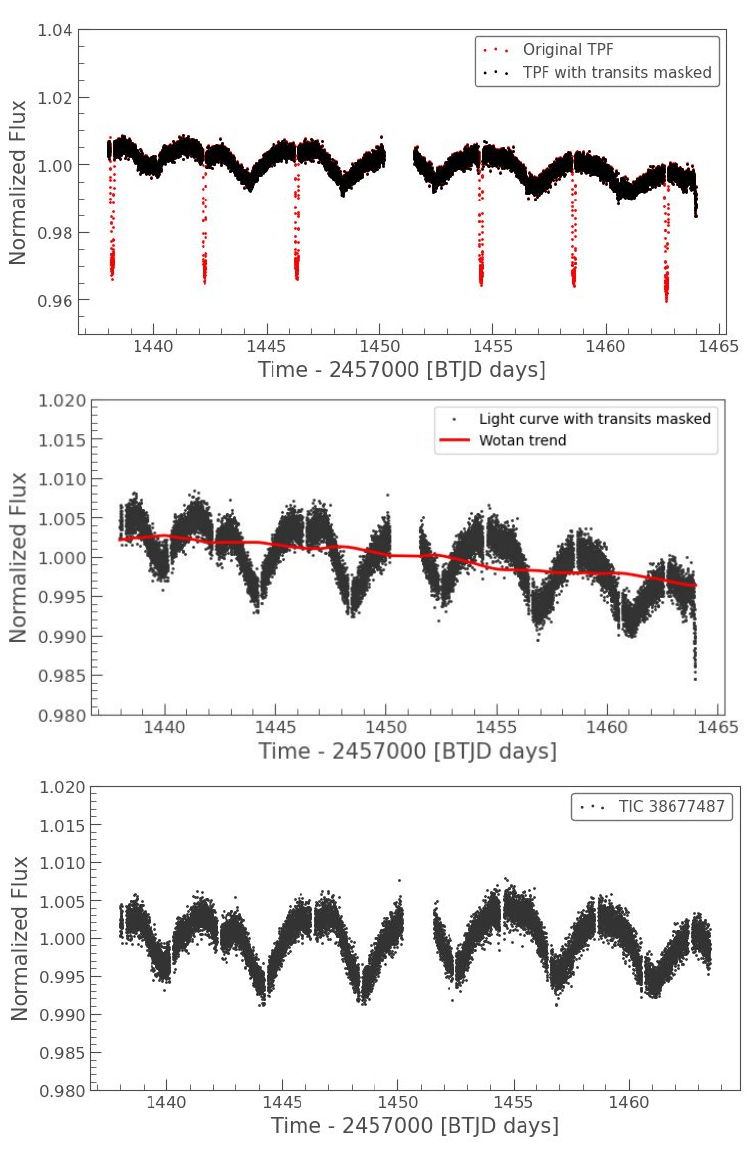}
    \caption{Demonstration of the procedure followed for preprocessing, taking an
example of EBLM J0351-07 (TIC 38677487). \textbf{Top: }SPOC corrected PDCSAP photometric light curve of Sector 5 for this target was downloaded, and the primary and secondary eclipses were cut. \textbf{Middle: }A \textsc{wotan} trend is fitted on the light curve with transits masked. \textbf{Bottom :}The final preprocessed light curve with instrumental systematics removed is obtained by subtracting the \textsc{wotan} trend and cutting the scattered data near 1464 days.}
    \label{fig:preprocessing}
\end{figure}

All these lightcurves were preprocessed to remove long-term instrumental systematics owing to the TESS orbital period. Multiple sectors of data were available for several targets, in which case the lightcurve corresponding to each sector was preprocessed individually and prepared for rotation period measurement with the same procedure as illustrated in Figure \ref{fig:preprocessing}. First, all the NaN values from the light curve were removed, and the flux was median normalized. Since eclipses are also periodic signals and can interfere with rotation period measurements, all the primary eclipses were cut, and the secondary eclipses were removed ad hoc. The light curve with eclipses removed may consist of variability that is astrophysical in origin and/or due to long-term instrumental artifacts. For this, the light curve is detrended by fitting and subtracting a \textsc{wotan} trend \citep{2019AJ....158..143H} with a sufficiently large window length of at least three times the orbital period. This helps remove only the long-term instrumental systematics while retaining rotation signatures and other astrophysical periodicities. Finally, a sharp change of flux near the start and end of the TESS orbit is observed in many sectors. Certain lightcurves also have some scattered data in between. We manually cut any such scattered data to obtain the final preprocessed lightcurves, ready to be used for rotation rate measurements.

\section{Methodology}\label{sec:methodology}
Starspots are less bright than the remaining stellar surface, introducing inhomogeneities on the star’s surface brightness. As the star rotates, it becomes brighter and fainter over its rotation period, and by estimating the periodicity in the lightcurve, we can find the stellar rotation period. However, one challenge is that the lightcurves are often flat, with no out-of-eclipse variability. This could be because - (i) the stellar activity is very high, and numerous spots are uniformly distributed all over the stellar surface, whose combined effect leads to a flat lightcurve. (ii) The star may have very few spots which could not be detected by TESS. (iii) There may be visible spots with long rotation periods ($> 14$ days) that we mistakenly subtracted as instrumental systematics. 

We adopted the following methodology for finding the periodicities in the lightcurves: 

\subsection{ACFs and Lomb-Scargle Periodograms}\label{sec:ACF}
The ACF method works by calculating the degree of similarity of a light curve with itself at a range of time lags. If the light curve consists of a periodic signal, we expect the ACF to produce peaks at lags corresponding to the signal's period and its integral multiples. While ACFs are not sensitive to the signal's phase and amplitude, they are robust and easy to automate. 

\begin{figure*}
    \centering
    \includegraphics[width = 0.9\textwidth]{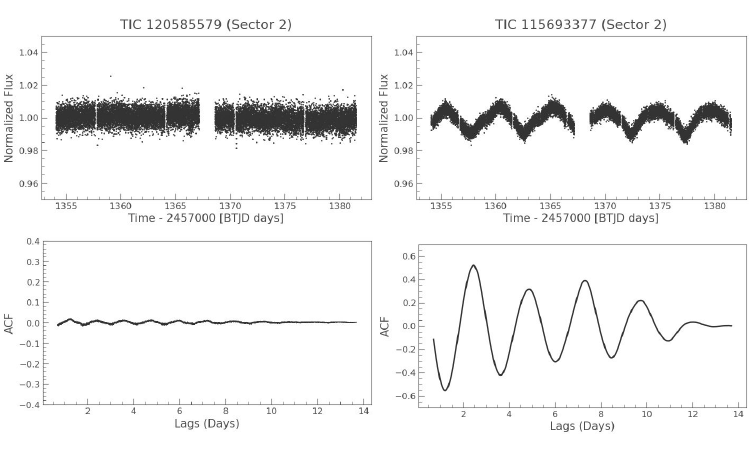}
    \caption{\textbf{Top:} Preprocessed lightcurve of EBLM J0017-38 (TIC 120585579)
with no out-of-eclipse variability (left). Preprocessed lightcurve of EBLM J0027-41 (TIC 115693377) exhibiting clear out-of-eclipse variability (right). \textbf{Bottom:} The corresponding ACF plots of the photometric data above them. The ACF plot at the left is just noise with no signal and corresponds to EBLM J0017-38, while the one on the right corresponds to EBLM J0027-41 and shows clear peaks. Only light curves with such out-of-eclipse variability were chosen for the ACF analysis}
    \label{fig:acfs}
\end{figure*}

To measure the periodicities in the TESS lightcurves in our sample, we first prepare a list of EBLMs exhibiting an out-of-eclipse variability by analyzing the preprocessed light curves by eye (refer to Figure \ref{fig:acfs}). Then, we plot ACFs individually for each sector of the eye-picked EBLMs using \textsc{astrobase.periodbase.macf} \citep{2013MNRAS.432.1203M,2014ApJS..211...24M} and used its built in error estimation for finding the error bars. The ACFs are smoothed using a Gaussian filter with a window size of 21 time lags. We select the highest ACF peak (first peak in most cases) as the initial ACF period of the light curve. This entire procedure was automated to save time and maintain consistency.

In addition to the ACF method, we also used Lomb-Scargle (LS) periodograms to evaluate the periodicities in the TESS lightcurves of the eye-picked EBLMs exhibiting an out-of-eclipse variability. The LS periodogram work by fitting sinusoids of different frequencies to the data using Fourier transform. By doing so, it recovers periodic trends in data and their corresponding periods \citep{1976Ap&SS..39..447L, 1982ApJ...263..835S}. Unlike the ACF method, periodograms can detect multiple periodicities in the data.

We generated the LS periodograms for the light curves exhibiting out-of-eclipse variability using \textsc{gatspy} python package \citep{2015ApJ...812...18V}. The period corresponding to the peak with the highest power was assigned as the initial LS period. All the peaks with power greater than 30\% of the highest peak's power were considered significant, and their corresponding periods were noted. 

\subsection{Two Sinusoid Fit}\label{sec:2sine}
Interestingly, $\sim 75\%$ of the initial ACF periods were roughly half the target's orbital period. The LS periodogram for $\sim 45\%$ of the EBLMs also had another significant peak at this period. Now, this could be a technical artifact or an astrophysical effect. Although the LS periodogram can detect multiple periodicities in the lightcurve, neither the ACF nor the LS periodogram gives information regarding the signal's phase or amplitude. Therefore, it is difficult to comment on the signal's source based on just these two methods. In order to verify whether this prevalent periodicity at approximately $1/2P_{\rm bin}$ in our sample has any astrophysical significance or not, we derived another method, called the "two sinusoid fit," which was previously introduced in \citet{Martin2024}. The two sinusoid fit was applied to our sample for measuring the stellar rotation periods, detecting two periodicities and identifying the source of these periodic signals.

\begin{figure*}
    \centering
    \includegraphics[width = \textwidth]{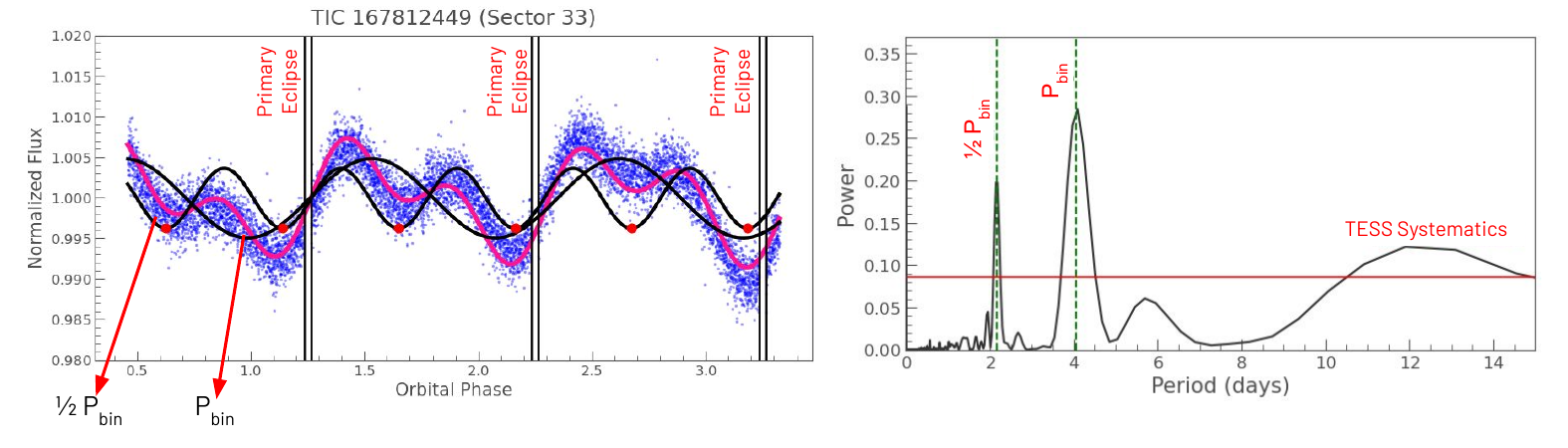}
    \caption{\textbf{Left:} Data segment representing one TESS orbit for Sector 33 of EBLM J0659-61 (TIC 167812449). The eclipses have been cut, and the orbital phase is set such that the primary eclipses occur at phase, $\phi$ = 0.25. The black sinusoids are the constituent sine waves of the overall fit represented by the magenta curve. The red dots correspond to the points where $1/2P_{\rm bin}$ sinusoid is minimized. These are plotted to check whether their phase coincides with that of the eclipses, which is one of the features of ellipsoidal variation. The mean of the phases represented by alternative red dots is taken as the phase at minima of the $1/2P_{\rm bin}$ sinusoid for this segment. \textbf{Right:} Lomb Scargle periodogram for the lightcurve shown on the left. The red line corresponds to 30\% the power of the highest peak, and all the peaks above this line are considered significant. The LS periodogram shows that the lightcurve has multiple periodicities, the stronger one at $P_{\rm bin}$ and the other at $1/2P_{\rm bin}$. There is an additional periodic signal that we attribute to TESS orbit systematics.}
    \label{fig:2sinefit}
\end{figure*}

First, the lightcurves of the EBLMs with an orbital period of more than three days were split into two 13.7 day long segments, each representing one TESS orbit. Within each segment, we fit two sinusoids using \textsc{scipy.optimize.curve\_fit}, taking $1/2P_{\rm RV}$ and $P_{\rm RV}$ as the initial period guesses. $P_{\rm RV}$ are the orbital periods of the EBLMs estimated through radial velocities in \citet{Triaud2017, Martin2021}. The initial guess values were chosen considering the findings from the ACF and periodogram analysis wherein a reasonably large number of EBLMs were found to be tidally locked and also exhibiting an additional periodicity at $\sim 1/2P_{\rm bin}$. The orbital phase between 0 to 1 is defined such that the primary eclipses lie at phase, $\phi$ = 0.25 as shown in Figure \ref{fig:2sinefit}. For tidally locked systems, the phase at the fitted sinusoid's minima indicates the location of active longitudes. Since each lightcurve segment will have multiple repeating sinusoidal minima, the mean of the phases at alternative minima of the $1/2 P_{\rm bin}$ sinusoid was noted. This gives us one phase for each lightcurve segment whose value would lie between 0 to 0.5 as we are looking at the $1/2 P_{\rm bin}$ signal. 

When the orbital period of an EBLM was less than three days, we first tried to fit sinusoids by splitting the data into multiple segments, each being at least four times the corresponding orbital period long. Only for the cases where we could not fit the sine curves due to insufficient data, we used an entire TESS orbit for the analysis. Splitting the data into these segments is essential to study rapid spot evolution.

The two-sinusoid fit method allows us to detect two periodic signals in the light curves and also gives information regarding the amplitude and phase of these signals. We attribute the prevalent periodicity at $1/2P_{\rm bin}$ in our sample to either of the two following effects:
\begin{enumerate}
    \item Ellipsoidal variation -  The gravitational interaction between stars in tight binary systems often leads to tidal deformation of the constituent stars' shape from a sphere to an ellipsoid. As a result,  we might observe periodic flux variations in the lightcurve of the binary system due to the change in the apparent cross-sectional area of the distorted stars as they orbit around each other. A signal originating from ellipsoidal variation has three features- (a) periodic at exactly $1/2P_{\rm bin}$, (b) phase, where the signal is minimized, coincides with the phase of the primary and secondary eclipses, i.e. phases = 0.25 and 0.75 in our two sinusoid fit analysis, and (c) signal's amplitude which remains relatively constant and can roughly be given by the following equation \citep{2010A&A...521L..59M}:
\begin{equation}
\label{eq:ellipsodal_amplitude}
A_{\rm ellip} \approx \alpha_{\rm ellip}\frac{M_{\rm B}}{M_{\rm A}}\left(\frac{R_{\rm A}}{a}\right)^3,
\end{equation}
where
\begin{equation}
\alpha_{\rm ellip} = 0.15\frac{(15 + u)(1+g)}{3-u}
\end{equation}
here $\alpha_{\rm ellip}$ can be calculated using the stellar gravity darkening coefficient, $g$ and the limb darkening coefficient, $u$ within the TESS bandpass.

The minimum of ellipsoidal variations can vary slightly from 0.25 if there is a lag in the tidal bulge (e.g. as seen in HS Hydrae, \citealt{Davenport2021}). We account for this in our classification scheme discussed in Sect.\ref{sec:classification}.
\\
\item Spots at active longitudes $180^\circ$ apart - If clusters of spots are located at two opposite longitudes separated by $180^\circ$ on the star, we may observe a periodic signal at $1/2P_{\rm bin}$. This was observed recently on the M-dwarf eclipsing binary, CM Draconis \citep{Martin2024}. In that case, the binary contains two equal brightness M dwarfs; hence, the active longitudes could have been on either or both stars. For our EBLM sample, we only see the light from the primary star; hence, we know it must be the host of the spots. Additionally, this phenomenon was observed in \textit{Kepler} Eclipsing Binaries by \citet{Lurie2017}, \textit{Kepler} main sequence stars by \citet{2014ApJS..211...24M}, and in single stars such as AU Mic \citet{2021A&A...649A.177M, 2021A&A...654A.159S}. This effect is also sometimes referred to as a BY Draconis variability which was first seen by \citet{Bopp1977, Bopp1981}. 

\begin{figure*}
        \centering
        \includegraphics[width = 0.99\textwidth]{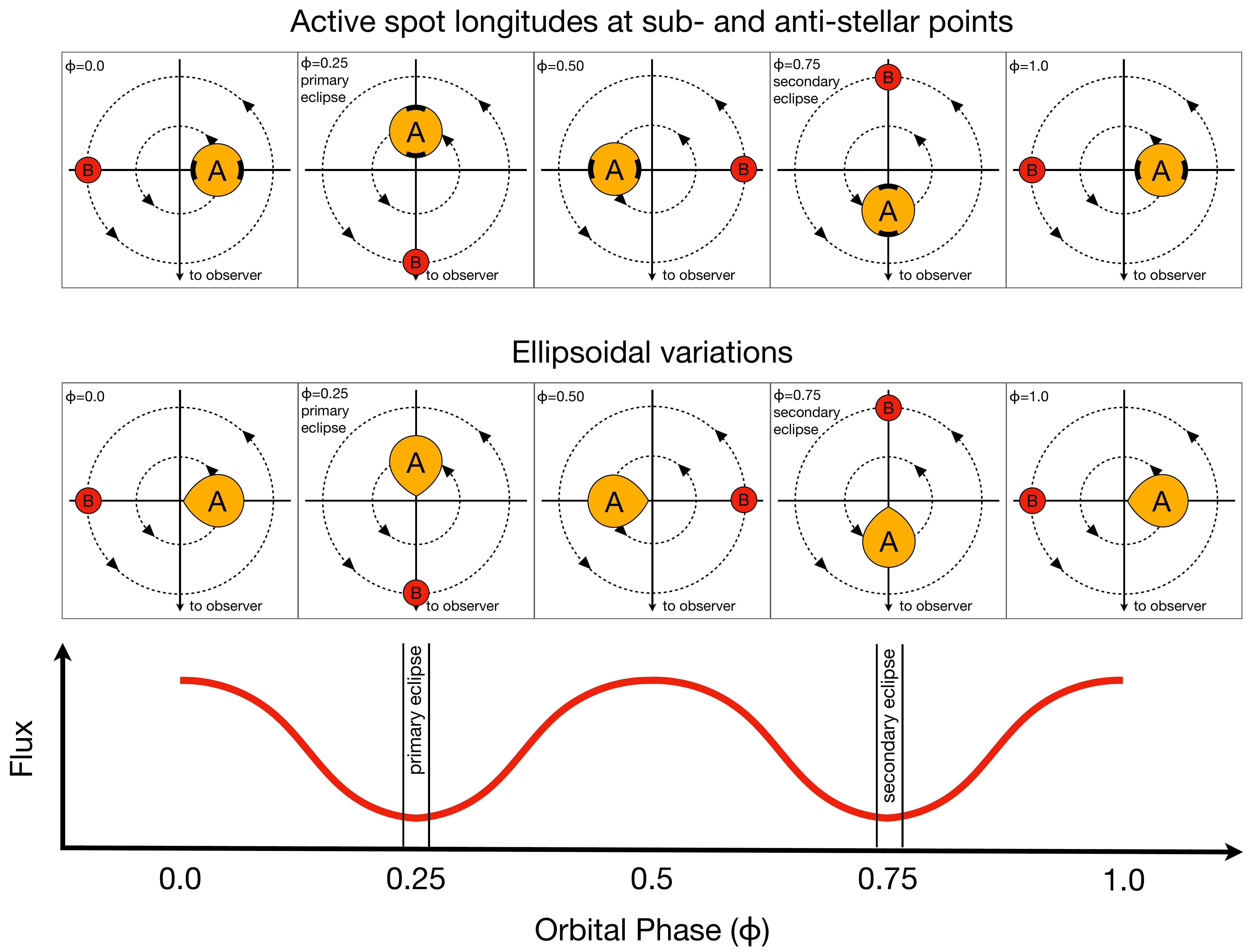}
        \caption{Illustration of two different physical effects with a similar effect on the lightcurve (${\bf bottom})$. The primary star (F/G/K) is ``A'' in yellow, and the secondary star (M-dwarf) is ``B'' in red. In both scenarios the two stars are tidally locked. {\bf Top:} an increased density of spots on the primary star at the sub-stellar point (closest point to the M-dwarf) and the anti-stellar point (farthest point to the M-dwarf). These active longitudes, illustrated by black regions on the star, make the star appear darker when facing the observer, which occurs near primary ($\phi=0.25$) and secondary ($\phi=0.75$) eclipses. {\bf Middle:} ellipsoidal variation due to a tidal distortion of the primary star, causing the asymmetric star to present a variable surface area and hence variable brightness to the observer. The star looks smallest and hence faintest near phase $\phi=0.25/0.75$. The two effects can be distinguished by spot evolution observed over a long timeseries, meaning that ellipsoidal variations are distinguished by being more stable. Active longitudes can also, a priori, be at any longitude (not just sub-/anti-stellar points).}
        \label{fig:active_vs_ellipsoidal}
\end{figure*}

We illustrate two different physical systems in Figure~\ref{fig:active_vs_ellipsoidal}, one with active longitudes located at sub- and anti-stellar regions and the other exhibiting ellipsoidal variation, but both having a similar photometric modulation. However, unlike ellipsoidal variations, which must always have a minimum at orbital phases of 0.25 and 0.75, these active longitudes are free to have minima at any orbital phase and not necessarily only at sub- and anti-stellar points. Hence, a signal from active longitudes that are $180^\circ$ apart may or may not have a minimum at 0.25 and 0.75. Furthermore, the evolution of the spots means that the phase and amplitude of the signal representing this effect can also change with time. We will see that for some of our binaries, the effect comes and goes between sectors.
\end{enumerate}

\subsection{Classification of Signals}\label{sec:classification}
 Only the signals originating from spot modulations can help us directly measure the star's rotation period. Although most ellipsoidal variables are tidally locked, i.e., the rotation period of a star in these systems is synchronized with its orbital period, the signals originating from them do not directly indicate the stellar rotation period. Therefore, the biggest challenge for our classification is distinguishing between ellipsoidal variation and spots at active, opposite longitudes. Ultimately when classifying signals, the main two characteristics of ellipsoidal variation are that its minimum phase must be at 0.25 (within our observational/fitting precision) and that this phase should be constant. The amplitude should also be constant, but since binaries may exhibit both ellipsoidal variations and spots, we do not use the amplitude in the criteria. The procedure used to classify the signals observed from the two sinusoid fit method into various categories based on their origin is described below.

\begin{figure}
         \centering
         \includegraphics[width = 1\linewidth]{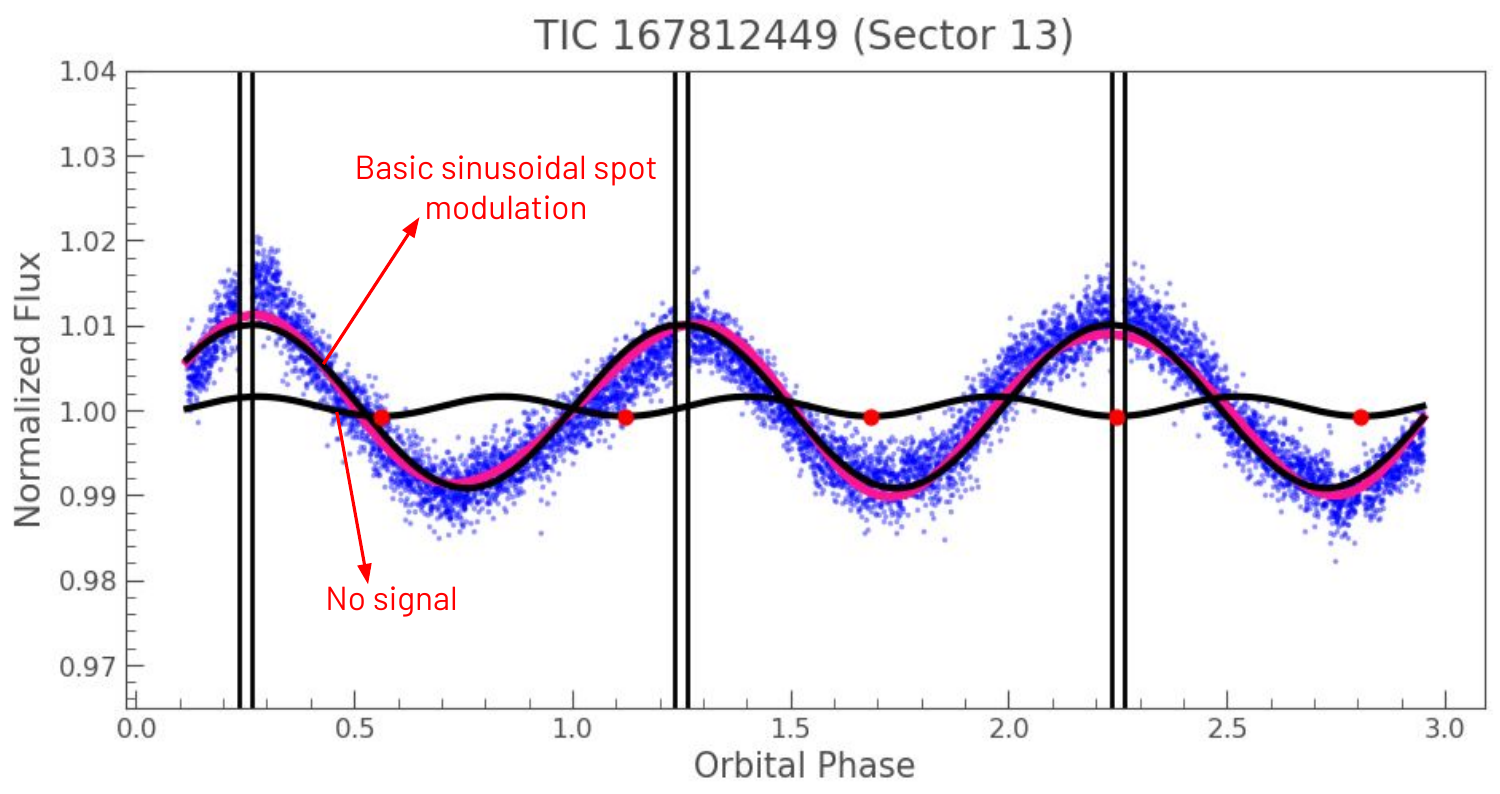}
         \caption{Source classification of signals derived from the two sinusoid fit analysis for a preprocessed data segment of EBLM J0659-61 (TIC 167812449). It is clear that the $1/2P_{\rm bin}$ sinusoid has a negligible amplitude and hence no astrophysical significance, which is why it is classified as "no signal". The $P_{\rm bin}$ sinusoid on the other hand has a significant ampltiude and hence classified as "basic sinusoidal spot modulation" which represents the rotation signature of the primary star.}
         \label{fig:classification_1}
\end{figure}

\begin{figure}
        \centering
        \includegraphics[width = 0.45\textwidth]{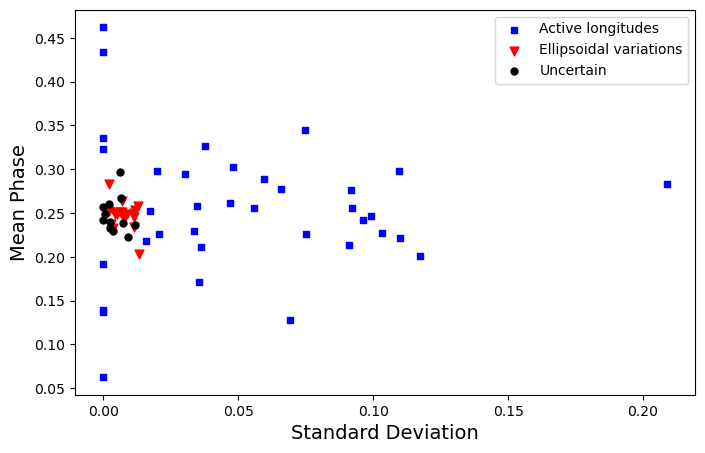}
        \caption{Classification of $1/2P_{\rm bin}$ sinusoidal signals as active longitudes or ellipsoidal variations. For each binary, we calculate the phase of the $1/2P_{\rm bin}$ signal minimum within each lightcurve segment (typically 13.7-day TESS orbits). Here we plot the mean phase against the standard deviation. Ellipsoidal variations should have a constant phase at 0.25 (we allow a range between 0.2 and 0.3 to be conservative). Active longitudes may have any phase between 0 and 0.5 and may vary. Red triangles are categorized as ellipsoidal variables. Blue squares are active longitudes. Black circles are uncertain because they have a phase near 0.25 but not enough data to tell if there is variation. Any data with a standard deviation of 0 has only 1 segment of data, and hence the standard deviation is not calculable.}
        \label{fig:half_binary_classification}
\end{figure}

\begin{figure}
    \centering
    \includegraphics[width = 0.49\textwidth]{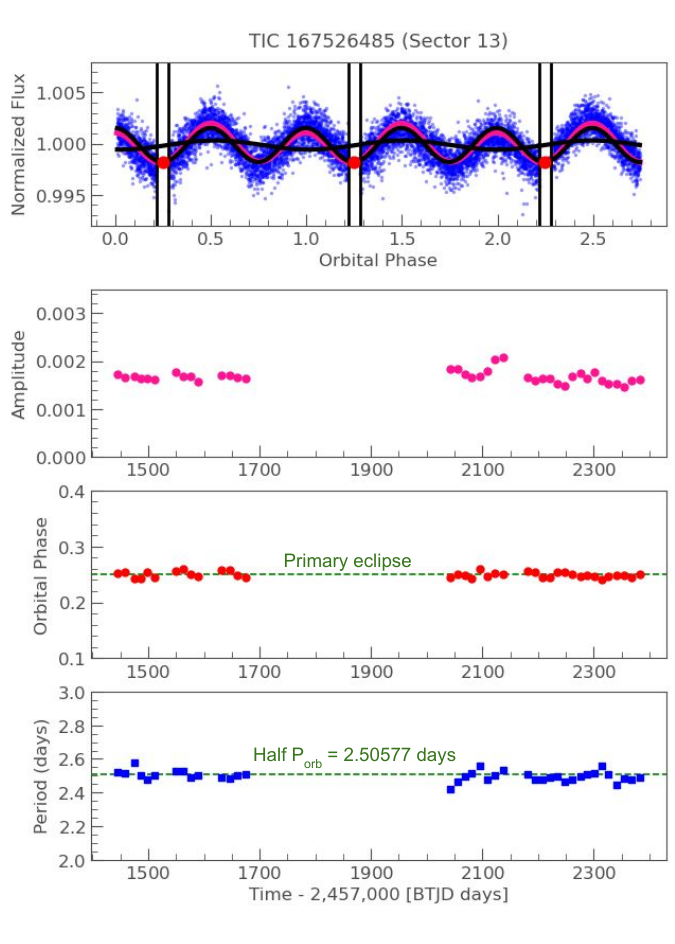}
    \caption{\textbf{Top: }Two sinusoid fit analysis of a preprocessed data segment of EBLM J0642-60 (TIC 167526485), clearly showing "no signal" for the $P_{\rm bin}$ sinusoid and "ellipsoidal variation" to be the source of $1/2P_{\rm bin}$ sinusoid. \textbf{Bottom: }The three plots at the bottom verify that the source of the $1/2P_{\rm bin}$ signal is ellipsoidal variation and depicts how its amplitude, phase at alternate minima, and period change over 38 segments of data (each equal to 1 TESS orbit) spread across 19 TESS sectors. It is clear that the amplitude remains roughly constant; its phase at alternate minima (red dots) coincides with that of the primary eclipses (phase $\approx 0.25$), and its period also remains constant and roughly equal to $1/2 P_{\rm RV}$. We therefore conclude that the source of the $1/2P_{\rm bin}$ sinusoid is ellipsoidal variation and not spots}
    \label{fig:EVs}   
\end{figure}

\begin{figure*}
     \centering
     \includegraphics[width = 0.99\textwidth]{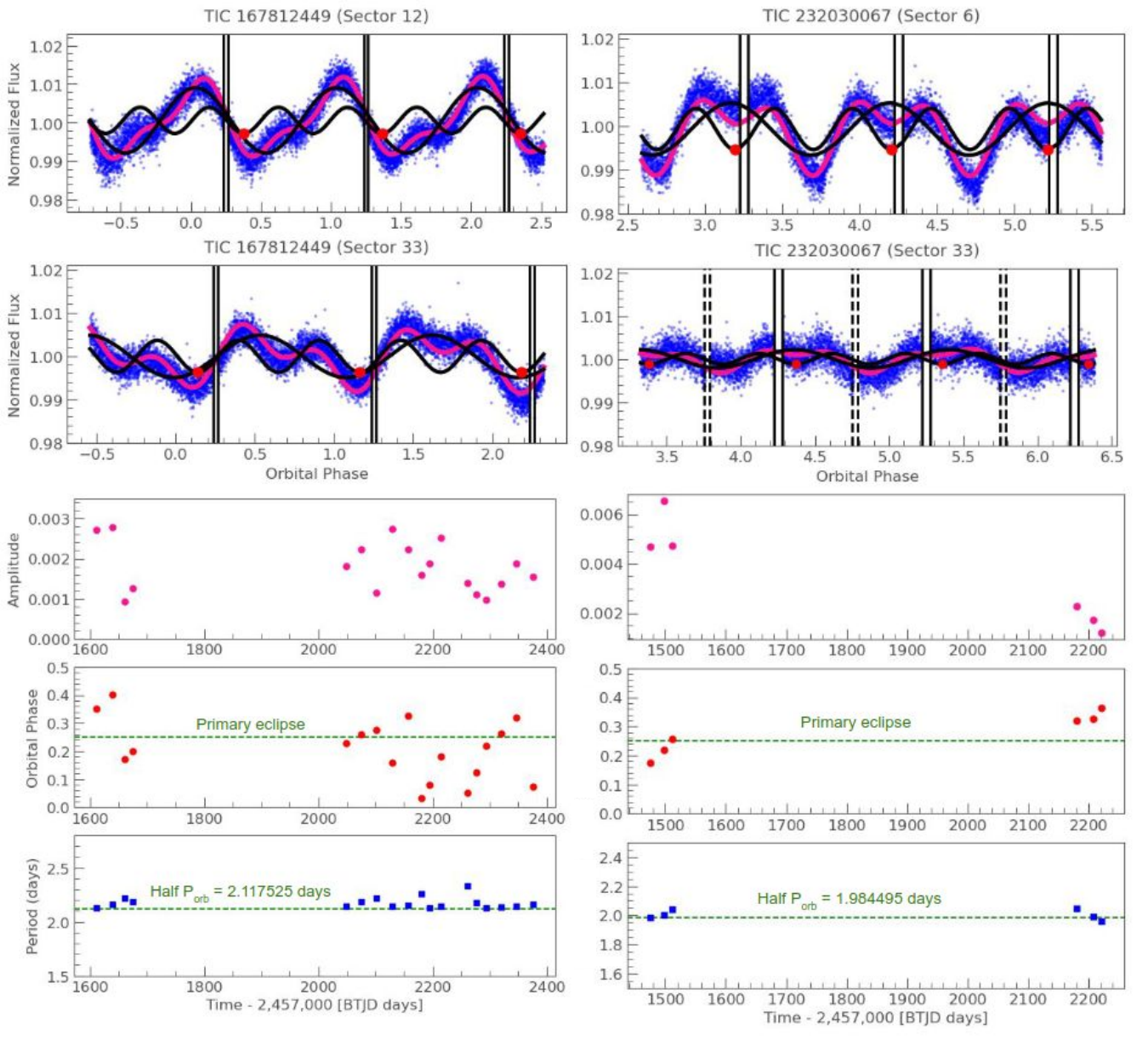}
     \caption{\textbf{Left : }Two sinusoid fit analysis of a preprocessed data segment of EBLM J0659-61 (TIC 167812449). Although the period of the $1/2P_{\rm bin}$ signal is pretty constant with a standard deviation of only 0.033, the orbital phase significantly varies across the sectors and has a standard deviation equals $\sim 0.11$. \textbf{Right :} Two sinusoid fit analysis of a preprocessed data segment of EBLM J0625-43 (TIC 232030067). Here again, we can see that the $1/2P_{\rm bin}$ period is very consistent with the $1/2P_{\rm RV}$, but the standard deviation is $\sim 0.052$. Since in both these examples, the orbital phase varies significantly (standard deviation $> 0.015$), we classify them as (d): spots at active longitudes that are $180^\circ$ apart in longitude and not ellipsoidal variations.}\label{fig:cm dra like effect}
 \end{figure*}

 \begin{enumerate}[label=(\alph*)]
     \item \textbf{No signal:} If the amplitude of either of the two sinusoids is too low, we conclude that they have no astrophysical significance. An example is shown in Figure \ref{fig:classification_1}, where there is a strong $P_{\rm bin}$ signal (indicative of spot modulation), but there is no significant $1/2P_{\rm bin}$ signal (i.e. no obvious ellipsoidal variations or active longitudes).
     \item \textbf{Sinusoidal spot modulation ($\mathbf{P_{\rm bin}}$ signal):} If the lower-frequency sinusoid, $P_{\rm bin}$ has a significant amplitude, we classify its source as spots and its period as the rotation period of the star. There may or may not be a significant $1/2P_{\rm bin}$ signal (in Figure \ref{fig:classification_1}, there is not). 
     \item  \textbf{Ellipsoidal variation  ($\mathbf{1/2 P_{\rm bin}}$ signal):} If there is a significant $1/2P_{\rm bin}$ signal, then we classify it as ellipsoidal variation if the mean minima phase across all the lightcurve segments is between 0.2 and 0.3, and its standard deviation is $< 0.015$. This is because the minima phase should coincide with the primary eclipse (phase 0.25), and the signal should be consistent, unlike spots that may vary. The range of 0.2 to 0.3 allows for both imprecise phase measurements and a possible offset due to a tidal bulge lag \citep{Davenport2021}. In Fig.~\ref{fig:half_binary_classification} we demonstrate the mean phase and standard deviation for all of our targets with a measured $1/2P_{\rm bin}$ signal. Targets for which we are confident that they show ellipsoidal variations are denoted by red triangles. An example of an ellipsoidal variable classification is shown in Fig \ref{fig:EVs}   
     \item \textbf{Spots at active longitudes ($\mathbf{1/2 P_{\rm bin}}$ signal):} All binaries which show a significant $1/2 P_{\rm bin}$ signal but do not fulfill the above criteria for ellipsoidal variation are classified as having active spot longitudes. In Fig.~\ref{fig:half_binary_classification} the active longitude signals are classified with blue squares. The active longitudes must be separated by $180^{\circ}$ in order to produce a signal at half the rotation period (here, equal to the orbital period). One example of active longitudes is shown in Figure \ref{fig:cm dra like effect} (right), where the active longitudes are roughly stable around phase 0.25. Another example is shown in Figure \ref{fig:cm dra like effect} (left), where the active longitudes change significantly in phase over the observations.
     \item \textbf{Uncertain ($\mathbf{1/2 P_{\rm bin}}$ signal):} There were a few cases where data for only one TESS sector was available, making it difficult to observe spot evolution. For such cases, we cannot say with certainty whether the source of the $1/2 P_{\rm bin}$ sinusoid is spots or ellipsoidal variation, and it could be either of them. These uncertain signals are classified as black circles in Fig.~\ref{fig:half_binary_classification}.

\end{enumerate}

Another thing to note here is that the classification of the $1/2 P_{\rm bin}$ signal as ellipsoidal variation or spots at opposite longitudes based on phase using the two sinusoid fit method is only valid for tidally locked systems and not otherwise. This is because, in this method, we define the orbital phase of the system and not the stellar rotational phase. We checked whether our EBLMs were tidally locked or not by using the ACF rotation periods ($P_{\rm ACF}$) given in Table \ref{rotation period catalog}, which were detected following the procedure given in Section \ref{sec:rotation_period_measurement}. Only if the value of $P_{\rm ACF}$ for a particular target was within the 10\% range of the corresponding $P_{\rm bin}$ the system was considered to be tidally locked. By default, the orbital phase of tidally locked systems happens to be equivalent to the star's rotational phase. This is not true for systems that are not tidally locked to each other, and hence we cannot and have not classified the source of the $1/2 P_{\rm bin}$ sinusoid for such cases.

We believe that we are conservative in our classification of stars as having active spot latitudes. It is possible that we classify some stars as ellipsoidal variables when they in fact host active longitudes that are both very close to the sub- and anti-stellar point and are very stable over our years of observation. \citet{Isik2007}, for example, demonstrated that highly active stars may have spot lifetimes considerably longer than on our own Sun. It is possible that with future observations from TESS or otherwise (e.g. PLATO) some binaries that are currently classified as ellipsoidal variables (or uncertain) will be re-classified as active longitudes.

\section{Results}\label{sec:results}

\subsection{Rotation Period Measurement and Validation}\label{sec:rotation_period_measurement}
We used a combination of three methods described in Section \ref{sec:methodology}- ACF, LS periodogram, and the two sinusoid fit for detecting the rotation periods of the EBLMs.

\begin{figure}
        \centering
        \includegraphics[width = 0.45\textwidth]{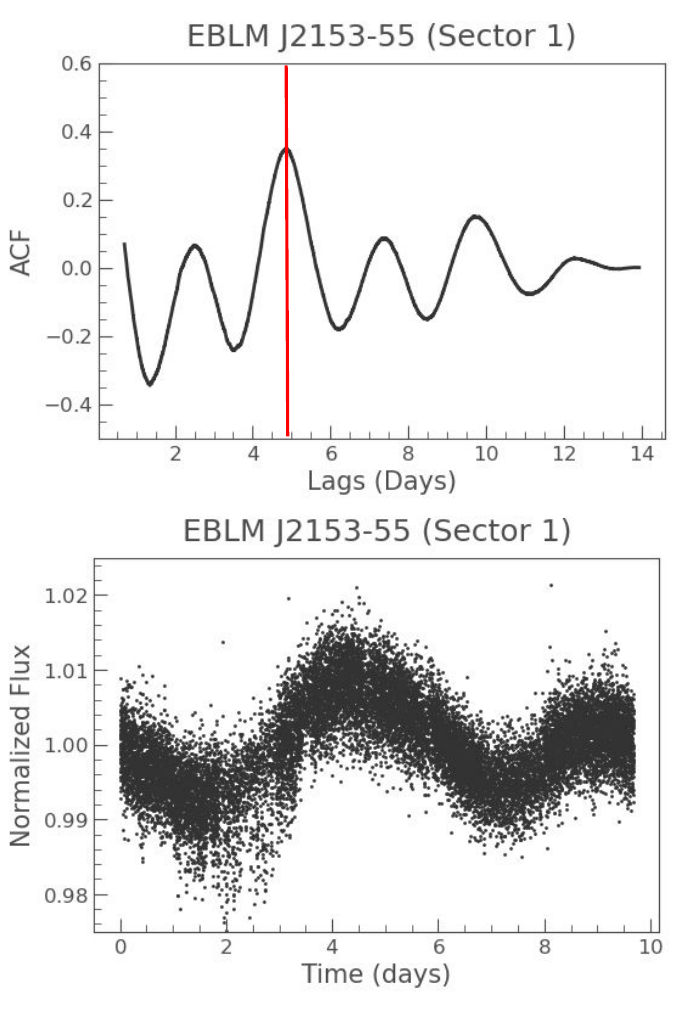}
        \caption{\textbf{ Top :} ACF plot of TIC 219322317 (EBLM J2153-55) showing that the initial ACF period corresponding to the highest peak is $\sim 5.07$ days. The orbital period of this target is $\sim 8.54$ days. Since the initial ACF period is less than 60\% the orbital period, and the $1/2P_{\rm bin}$ signal is classified as NA from the two sinusoid fit, we double the initial ACF period and phase fold it at this value to see if there is a strong signal. \textbf{Bottom :} It is evident that the phase folding indeed yields a strong signal, and the doubled period ($\sim 10.14$ days)is thus taken as the final ACF period }
        \label{fig:ACF phase folding}
\end{figure}

 As mentioned earlier, $\sim 75\%$ of the initial ACF periods were approximately equal to $1/2P_{\rm RV}$, creating a confusion whether their source is an ACF artifact or astrophysical effects like spots at opposite longitudes and ellipsoidal variation. The two sinusoid fit method solved this problem by classifying the sources of all the observed signals which were subsequently used for our rotation period estimations. First, the EBLMs with the initial ACF period of less than 60\% the system's orbital period were identified. To verify the ACF results, we searched for a corresponding significant peak in the LS periodogram for these systems. An additional significant peak was indeed observed in $\sim 60\%$ of the identified EBLMs. We then checked the source of the $1/2P_{\rm bin}$ signal from the two sinusoid fit method for these identified EBLMs. If the source of the $1/2P_{\rm bin}$ sinusoid was-
    
\begin{enumerate}
    \item Ellipsoidal variation/ uncertain (either spots at opposite longitudes or ellipsoidal variation): We do not find the ACF rotation period for that EBLM. 
    \item Spots at opposite longitudes/ NA (the systems whose $1/2P_{\rm bin}$ signal classification was not possible as they are not tidally locked): We double the initial ACF period and verify it by phase folding the corresponding lightcurve at this period. Doubled periods that give a clear signal upon phase folding are noted down as the stellar rotation period (Refer to Figure \ref{fig:ACF phase folding}). 
    \item No signal: For this case, we phase fold the corresponding light curve at twice the initial ACF period and check for a rotation signature. If the phase folding yields no signal, we do not include an ACF rotation period in our catalog. However, if it yields a signal, we take twice the initial ACF period as the rotation period and conclude that the initial ACF period was probably an ACF artifact. 
\end{enumerate}    
For the remaining cases, we phase fold the corresponding lightcurves at the initial ACF periods to verify them. All the periods which do not give a signal upon phase folding are discarded and the remaining are noted down as the star's rotation period. The final ACF rotation period of an EBLM is taken as the mean ACF rotation period across all TESS sectors studied for that particular EBLM. 

In the case of LS periodograms, the highest peak is chosen as the initial LS period, and then the same procedure used for finding the ACF rotation periods is followed.

Both the ACF rotation periods ($P_{\rm ACF}$) and the LS rotation periods ($P_{\rm LS}$) along with the periods derived from the two sinusoid fit method ($1/2P_{\rm bin}, P_{\rm bin}$) and the sources (no signal, rotation signature, ellipsoidal variations, spots at opposite longitudes or uncertain) of their signals are cataloged in Table \ref{rotation period catalog1}. We also give a nominal rotation period in our catalog, which is nothing but the $P_{\rm ACF}$. If $P_{\rm ACF}$ could not be estimated, the nominal period is taken as $P_{\rm LS}$. Finally, for the cases where both $P_{\rm ACF}$, and $P_{\rm LS}$ could not be estimated, but the source of the $P_{\rm bin}$ signal is classified as sinusoidal spot modulation, the period corresponding to the $P_{\rm bin}$ sinusoid is taken as the nominal rotation period and its error is defined as the standard deviation of the period values measured across all the TESS orbits. We have successfully cataloged the nominal rotation periods of 69 EBLMs and discovered ellipsoidal variations in 17.


\subsection{Active Spot Longitudes in Tight Binaries}

Out of our 81 binaries we determine that 41 show evidence of active spot longitudes. Recall that we only see light (and hence spots) on the primary FGK star, as opposed to the secondary M-dwarf star (even though the M-dwarf is probably also spotted). We are also only sensitive to active longitudes if they are roughly $180^{\circ}$ apart in latitude, since that is what produces a $1/2P_{\rm bin}$ signal in tidally locked binaries. This is not to say that active longitudes with different separations are impossible, but rather they would produce a more complex signal that our simple spot detection methodology is not able to detect and characterise.

\begin{figure*}
     \centering
     \includegraphics[width = 0.99\textwidth]{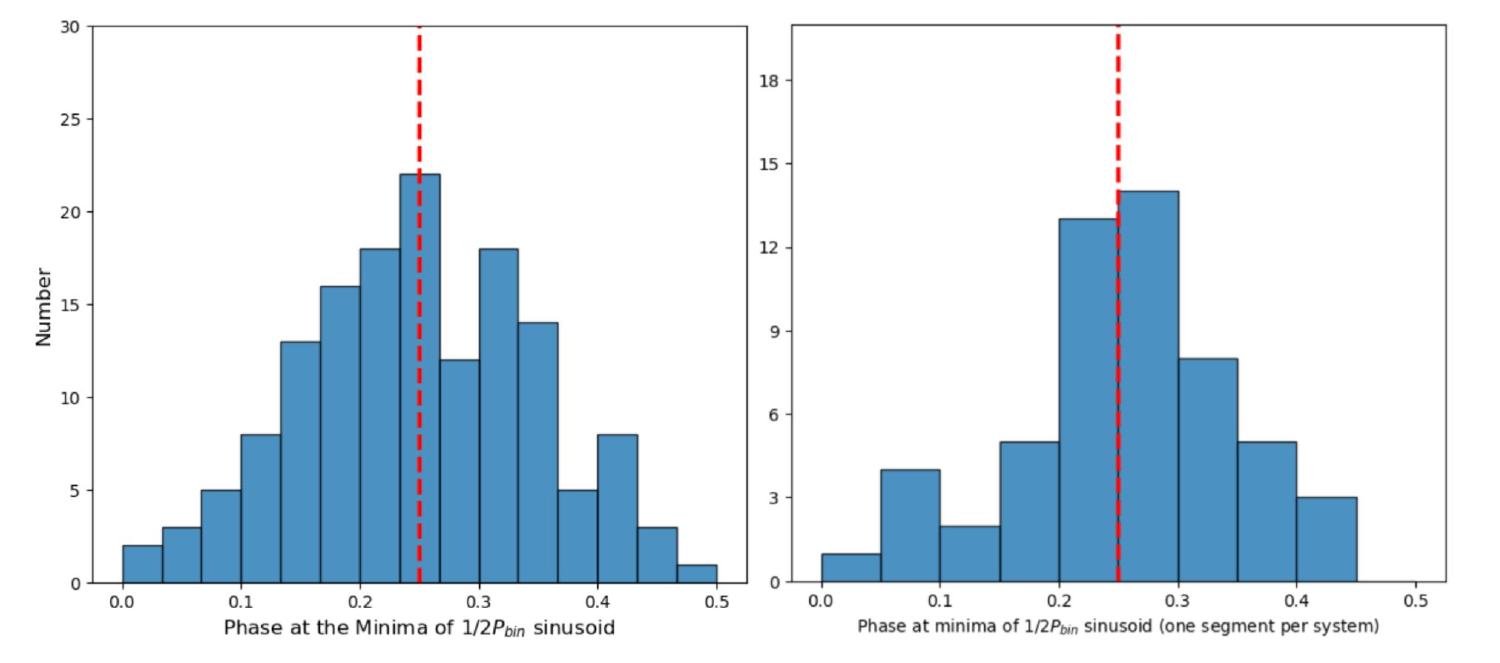}
     \caption{\textbf{Left : } Histogram of the minimum phase for all binaries showing evidence of active longitudes (a significant signal at $1/2P_{\rm bin}$ but with ellipsoidal variation ruled out). The phase is defined relative to the orbital phase, which is why it is only defined between 0 and 0.5, where 0.25 corresponds to a primary eclipse. For each binary the phase is calculated in every light curve segment, and hence each binary will add a different count to the histogram depending on how much it was observed by TESS. Right : Histogram of the minimum phase for all binaries showing evidence of active longitudes, but this time, phase from only one light curve segment for each binary adds to the histogram count. }\label{fig:active_longitudes_histogram}
 \end{figure*}

In Fig.~\ref{fig:active_longitudes_histogram} \textbf{(left)} we show a histogram of the minimum phase of the $1/2P_{\rm bin}$ signal for all binaries showing active longitudes. A point is added to the histogram for every lightcurve segment per target. That means that binaries observed more often will contribute to the histogram more, but it also accounts for the variation in phase due to spot evolution. All binaries where we either suspect or are confident of ellipsoidal variation are excluded. To make sure that our results are not biased by few stars with many observed segments and very stable active longitudes, we also show a histogram in Fig. \ref{fig:active_longitudes_histogram} (Right) where a point is added for only one lightcurve segment (which is randomly chosen) per target.

We see in Fig.~\ref{fig:active_longitudes_histogram} a strong preference for phase minima near an orbital phase of 0.25, corresponding to the primary eclipse. Since the period is half the binary period, there will be also an equivalent minima near 0.75 (secondary eclipse). Recall that a priori the active longitudes could be at any longitude on the primary star, and hence this phase minimum could be anywhere from 0 to 0.5 (i.e., a flat histogram). The fact that we have a concentration at phase 0.25 corresponds to a preference for active longitudes at the sub- and anti-stellar points. Otherwise said, stars in eclipsing binaries tend to have a pile-up of spots facing directly towards and away from the other star. One thing to note here is that this result is not  simply recovering the central limit theorem. If we histogrammed a mean phase per star then the central limit theorem would dictate that the histogram be approximately normal, with a mean at 0.25, even if the underlying distribution of phases were uniform between 0 and 0.5. We do not do this. Rather, our histograms correspond to either all lightcurve segments per target (Fig. \ref{fig:active_longitudes_histogram} (left)) or  one random lightcurve segment per target (Fig. \ref{fig:active_longitudes_histogram} (right).


It is possible that the true preference for active longitudes at phase 0.25 is stronger than indicated by Fig.~\ref{fig:active_longitudes_histogram}. Consider Fig.~\ref{fig:half_binary_classification}, used in the classification of the $1/2P_{\rm bin}$ signals. By construction/expectation, all of the ellipsoidal variation signals have a phase near 0.25 and a small standard deviation (i.e. the phase is nearly constant). It is possible that some of these are in fact active longitudes but with spots that evolve on a much longer timescale. There are also 12 binaries with an uncertain signal (black circle in Fig.~\ref{fig:half_binary_classification}). They are uncertain because the phase is near 0.25 but there is not enough data to see if the signal varies. Such signals are excluded from Fig.~\ref{fig:active_longitudes_histogram}, and their inclusion would increase the peak at phase 0.25.

On the other hand, it is possible that some ellipsoidal variables were mistakenly classified as active longitudes, and hence are wrongly contributing to Fig.~\ref{fig:active_longitudes_histogram}. Even though we require at least two TESS sectors, it's possible that the data quality is low and hence our phase measurement, which is the main discriminant, is faulty. Revisiting these data when more TESS observations become available will help alleviate this problem. Ultimately, if Fig.~\ref{fig:active_longitudes_histogram} were overwhelmingly contaminated by ellipsoidal variable signals, we would expect to see more sharply peaked function at a phase of 0.25, and not the observed smooth distribution.

An earlier study by \citet{Olah2006} on 12 binaries with active longitudes found that three of the sample had active longitudes at both sub- and anti-stellar points. The rest showed active longitudes at either quadrature ($90^{\circ}$ off sub- and anti-stellar points), or both sub-stellar points and other longitudes, or a variety of longitudes. Interestingly, their three binaries with sub- and anti-stellar active longitudes were the binaries with the most unequal mass ratio ($q=m_{\rm B}/m_{\rm A} < 0.6$). The entire EBLM, by its construction, has unequal mass ratios ($q <0.5$). However, their $P_{\rm orb}$ is 4.768,
24.428, and 24.649 whereas the $P_{\rm orb}$ of all the EBLMs given in Table \ref{rotation period catalog1} is less than 10 days. Another difference is that the \citet{Olah2006} sample contains giant and sub-giant stars, whereas the primaries in the EBLM sample are typically on the main sequence, so the two samples may not be comparable.

\subsection{Spot Migration}\label{subsec:migration}
\begin{figure*}
     \centering
     \includegraphics[width = 0.99\textwidth]{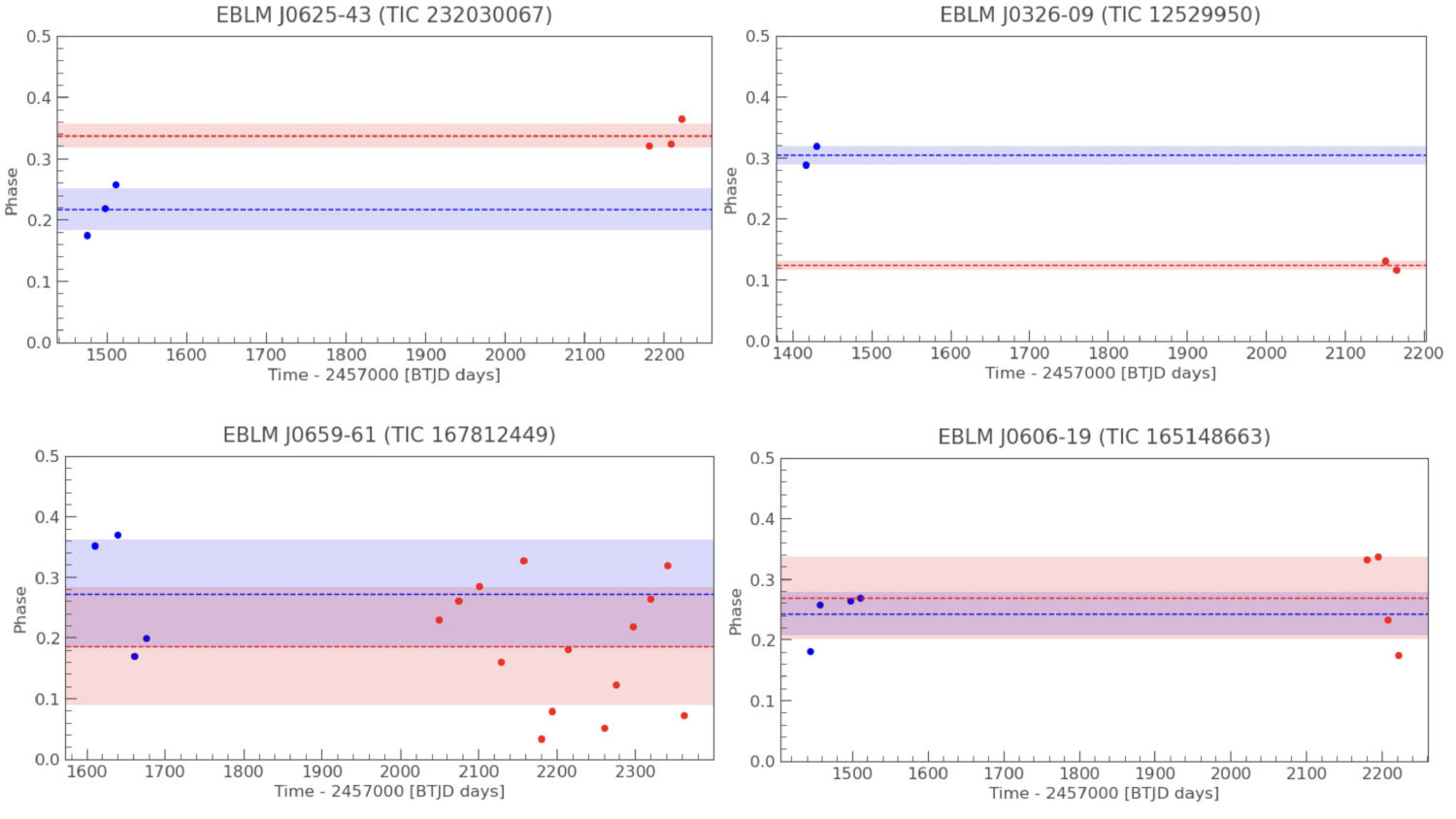}
     \caption{Few examples to see if there is any longitudinal spot migration in some selected EBLMs or not. The blue and red dots correspond to the minimum phase for the TESS orbits lying in the first and second half of TESS data respectively. The blue and the red dotted lines represent the respective mean minimum phase for the two halves, and the shaded area represents the area spanned by their respective standard deviations. The \textbf{Top} panel shows 2 out of 4 examples of cases where the two dotted lines of the mean minimum phase for the two TESS halves is separated by atleast twice their respective standard deviations i.e. they have a 2-sigma spot migration. The \textbf{bottom} panel shows examples for the cases where the shaded areas overlap and there is no detectable longitudinal spot migration.}\label{fig:spot migration}
 \end{figure*}  
On the Sun, the spots migrate over the 11-year solar activity cycle. They typically appear at latitudes of roughly $\pm 30^{\circ}$ at the start of the cycle, and then proceed to migrate towards the equator and disappear. This gives rise to the famous ``butterfly diagram''. We investigate a similar longitudinal and latitudinal  spot migration in our sample of EBLMs. 

Longitudinal spot migration may effect the observed period. If the spot migrates in the direction of rotation then the observed rotation period will be shorter than reality, and vice versa. This is akin to the sidereal vs synodic period of lunar phases. However, if this longitudinal migration is at a constant speed then the observed rotation period will be constant, albeit slightly different to the true rotation period. Longitudinal spot migration will also affect the observed spot phase. This latter effect is more visible in our sample, e.g. Fig.~\ref{fig:spot migration}.

Latitudinal spot migration may also affect the observed period, as different latitudes of the star rotate at different speeds owing to differential rotation. Latitudinal migration of spots may affect the observed spot phase solely becuase we calculate phases relative to the fixed orbital period. This orbital period will be close to the rotational period, due to tidal locking, but there may be subtle differences owing to differential rotation. If the spot period changes, then the spot phase will also change since we always phase-fold relative to the orbital period.

Overall, we expect longitudinal spot migration to cause a change in spot phase but not necessarily the observed rotation period. Latitudinal spot migration, on the other hand, should result in both changes to the spot period and spot phase, if there is differential rotation.

To check if there is any observational evidence of multiple rotation period signals due to differential rotation, we go back to our LS periodogram analysis described in \ref{sec:ACF} and looked at the additional significant peaks. All these peaks were attributed to either spots at opposite longitudes or ellipsoidal variation and not differential rotation. We also observed that the period corresponding to the $P_{\rm bin}$ sinusoid, measured across different TESS orbits using the two sinusoid fit method, is very consistent, whereas if the spots were migrating to different latitudes, we would have seen a change in this period due to differential rotation. We therefore conclude that there is no evidence of differential rotation and hence latitudinal spot migration in our sample. 

This could be because we are dealing with F/G/K primaries, and just like Sun these stars are more likely to have spots distributed within $\pm 30^\circ$ from the equator, meaning they have less differential rotation. Additionally, studies have also shown that differential rotation is less pronounced in faster rotating stars \citep{1998ApJR, 2003A&Aschmitt, 2007collier, Lurie2017}, and the detected rotation periods for more than $~72\%$ of the EBLMs in Table \ref{rotation period catalog1}, is less than 5 days and nearly all the EBLMs under study have a rotation period of less than 12 days.  Also, the targets for which we did the spot migration analysis, have an approximate dataset length of only $\sim 700 - 800$ days (about 2 years). However, the length of the mean cycle is found to be 9.5 yr for F-type stars, 6.7 yr for G-type stars, 8.5 yr for K-type star. \citep{2016A&A...595A..12S}. Henceforth, it is possible that one would not expect to see much latitudinal spot migration (and differential rotation) with this length of the dataset. 

However, we do have preliminary evidence of longitudinal spot migration that can be seen in Fig. \ref{fig:cm dra like effect} (Right) where spots migrate from phase $\approx 0.2$ to $\approx 0.35$ in about 700 days. To investigate this further, we shortlist the binaries where we detect active longitudes and that have atleast one TESS sector of data between 1-15 TESS Sectors, which we call the first half, and one TESS sector of data in the second half i.e. TESS data beyond sector 15. As a rough metric, we calculate the mean and standard deviation of the phase at the minima of the $1/2 P_{\rm bin}$ sinusoid corresponding to each TESS orbit in the first and the second half of the TESS data. To see if there is any observable longitudinal spot migration, we check whether standard deviations around the mean minimum phase of the two halves overlap or not. Out of 18 systems, where we check this, four have 2-sigma migration as seen in Fig. \ref{fig:spot migration} (Top), seven have 1-sigma migration, and the standard deviations around the mean minimum phase for the remaining 11 binaries overlap for the two TESS halves as seen in Fig.\ref{fig:spot migration} (bottom) and hence they have no detectable longitudinal spot migration. 

One effect we might be sensitive to is finite spot lifetimes. From Kepler data, spots were found to last between 10 and 350 days \citep{Namekata2019}, and for practically all of our targets our timespan is longer than one year. If the spots appear and disappear at the same location then we will not see much of a signal, but if they do so at different latitudes and longitudes then this could introduce a phase-variation in our fitted sinusoids.

Overall, we believe our sample shows some evidence for longitudinal spot migration but the consistency of the observed rotation periods argues against latitudinal spot migration connected with differential rotation. An improved understanding could come from both new data in upcoming TESS cycles and a re-analysis using a more sophisticated lightcurve modelling tool such as \textsc{PHOEBE} \citealt{Prsa2005}.

\section{Conclusion} \label{sec:conclusion}
We study modulations in the TESS photometry of 162 binaries in the EBLM (Eclipsing Binary Low Mass) catalog. All of them contain one bright F/G/K star and one faint M-dwarf secondary. In this sample 81 show detectable signs of modulation, 69 of which are consistent with star spots on the primary star This provides a measurement of its rotation rate, indicating that the majority are tidally locked, as expected for a sample of tight binaries. In roughly half of these tidally locked binaries the periodicity of the spot modulation, in particular a signal at half the binary's orbital period, is indicative of two clusters of spots separated by $180^{\circ}$ in longitude. These so-called active longitudes are not distributed at random on the primary stars, but rather have a preference for the sub- and anti-stellar points. Otherwise put, we have evidence for a connection between stellar activity and binarity, at least for close, tidally locked systems.

In our analysis we also detect a strong ellipsoidal variation on 17 binaries. Ellipsoidal variations also produce a periodicity at $1/2P_{\rm bin}$. Whilst this could be confused for the aforementioned effect of active longitudes (Fig~\ref{fig:active_vs_ellipsoidal}), we distinguish between the two effects by the phase and consistency of the $1/2P_{\rm bin}$ signal. We also identify systems where we cannot make a confident classification either way. bf We find evidence for longitudinal spot migration in some systems, owing to a change in the spot phases. However, the observed rotation periods remain constant, and hence with our current data precision we do not see evidence for latitudinal spot migration and differential rotation.

Our work has implications for theoretical studies of the magnetic interactions of close binaries. It also has implications for observational studies of eclipsing binaries in the hunt for ultra-precise fundamental stellar parameters (including mass and radius). Indeed, such measurements are a primary goal of the EBLM survey. The most reliable measurement of an M-dwarf's radius is in an eclipsing binary. If the primary star's spot distribution is not random but concentrated near the sub-stellar point (i.e. near the primary eclipse) then this might bias the measured radii. This might be relevant for the so-called ``radius inflation problem'', where radius measurements of M-dwarfs are higher than expected when compared with models. Quantifying this effect is left to a future work. Such a study might also benefit from a more sophisticated spot and ellipsoidal variation model (e.g. \textsc{PHOEBE}, \citealt{Prsa2005}).
 
\section*{Data availability}
All data used in this analysis (e.g. light curves) can be provided upon reasonable request to the authors.

\section*{Acknowledgements}

Support for this work was provided by NASA through the NASA Hubble Fellowship grant HF2-51464 awarded by the Space Telescope Science Institute, which is operated by the Association of Universities for Research in Astronomy, Inc., for NASA, under contract NAS5-26555. This work was based on photometry taken by the TESS spacecraft. In particular, we benefited from short cadence 120 second photometry, courtesy of Guest Investigator programs led by the EBLM team. We specifically thank Alison Duck for her efforts with the writing of the Cycle 5 and 6 proposals. We also thank Pierre Maxted, Amaury Triaud and the entire EBLM collaboration for their tireless work creating this beautiful sample of low mass eclipsing binaries.

\bibliographystyle{mnras}
\bibliography{references}

\appendix

\begin{table*} \label{table 1}
\caption{Rotation rate measurements and classification of ellipsoidal variations and active longitudes in 81 EBLM binaries. The orbital periods ($P_{\rm RV}$) are taken from radial velocities in \citet{Triaud2017,Martin2021}. $P_{\rm ACF}$ and $P_{\rm LS}$ are the photometric modulation periods derived by the autocorrelation function and Lomb Scargle periodogram, respectively. The two rightmost columns are the periods derived from a two sinusoid fit. The source is (a) no signal detected, (b) basic spot modulation, (c) ellipsoidal variation and (d) spot modulation from active longitudes $180^{\circ}$ apart. These different sources are described in Sect.~\ref{sec:classification}. As we cannot comment on the source of the $1/2P_{\rm bin}$ signal for systems that are not tidally locked, it is defined as NA for those cases.}
\label{rotation period catalog1}
    \begin{tabular}{|c|c|c|c|cc|cc|c|c|}
    \hline
        EBLM Name & $P_{\rm RV}$ & $P_{\rm ACF}$ & $P_{\rm LS}$ & $1/2 P_{\rm bin}$ & Source & $P_{\rm bin}$ & Source & No. of TESS & Nominal $P_{\rm rot}$\\ 
        & (days)&(days)&(days)&(days)& &(days)& &Sectors & (days) \\\hline
        EBLM J0013-31 & 3.4373 & - & - & 1.725 & (c) & - & (a) & 2 & - \\ \hline
        EBLM J0021-16 & 5.96727 & 7.042±0.061 & 6.8833 & 3.454 & NA & 6.551 & (b) & 1 & 7.042±0.061 \\ \hline
        EBLM J0027-41 & 4.92799 & 4.926±0.158 & 4.8518 & 2.526 & (d) & 4.908 & (b) & 1 & 4.926±0.158 \\ \hline
        EBLM J0035-69 & 8.4146 & 8.226±0.093 & 8.1982 & 4.261 & (d) & 8.225 & (b) & 2 & 8.226±0.093 \\ \hline
        EBLM J0040+00 & 7.231 & 11.728±0.134 & - & - & - & - & - & 1 & 11.728±0.134 \\ \hline
        EBLM J0048-66 & 6.64927 & 6.656±0.001 & - & - & (a) & 6.593 & (b) & 2 & 6.656±0.001 \\ \hline
        EBLM J0057-19 & 4.30051 & 4.891±0.079 & 4.985 & 2.624 & (d) & 4.947 & (b) & 2 & 4.891±0.079 \\ \hline
        EBLM J0157-11 & 3.88745 & 4.26±0.111 & 4.4715 & 2.112 & NA & 4.432 & (b) & 2 & 4.26±0.111 \\ \hline
        EBLM J0239-20 & 2.77868 & 2.853±0.121 & 2.9587 & 1.44 & (d) & 2.905 & (b) & 2 & 2.853±0.121 \\ \hline
        EBLM J0247-51 & 4.00785 & 4.044±0.061 & 4.1512 & 2.066 & (d) & 4.1 & (b) & 4 & 4.044±0.061 \\ \hline
        EBLM J0326-09 & 2.4004 & 2.401±0.05 & 2.4611 & 1.209 & (d) & 2.486 & (b) & 2 & 2.401±0.05 \\ \hline
        EBLM J0339+03 & 3.58067 & 4±0.024 & - & 1.952 & (d) & 3.458 & (b) & 1 & 4±0.024 \\ \hline
        EBLM J0345-10 & 6.06135 & 6.251±0.033 & - & 3.133 & (d) & 6.036 & (b) & 1 & 6.251±0.033 \\ \hline
        EBLM J0351-07 & 4.0809 & 4.122±0.052 & 4.3868 & 2.169 & (d) & 4.397 & (b) & 2 & 4.122±0.052 \\ \hline
        EBLM J0356-18 & 1.30211 & 1.334±0.025 & 1.29 & 0.648 & (d) & 1.337 & (b) & 3 & 1.334±0.025 \\ \hline
        EBLM J0400-51 & 2.69208 & 2.686±0.036 & 2.6404 & 1.353 & (d) & 2.673 & (b) & 4 & 2.686±0.036 \\ \hline
        EBLM J0432-33 & 5.30549 & 5.297±0.005 & 5.2932 & 2.659 & (d) & 5.474 & (b) & 4 & 5.297±0.005 \\ \hline
        EBLM J0440-48 & 2.54304 & 2.555±0.062 & 2.5936 & 1.298 & (d) & 2.58 & (b) & 3 & 2.555±0.062 \\ \hline
        EBLM J0443-06 & 3.11192 & - & - & 1.53 & (c) & 3.101 & (b) & 2 & 3.101±0.076 \\ \hline
        EBLM J0453+06 & 0.80415 & 0.804±0.006 & 0.8026 & 0.4 & (d) & 0.803 & (b) & 2 & 0.804±0.006 \\ \hline
        EBLM J0454-09 & 5.01345 & 4.932±0.066 & 4.8775 & 2.409 & (d) & 4.854 & (b) & 2 & 4.932±0.066 \\ \hline
        EBLM J0502-38 & 3.2563 & 3.263±0.097 & 3.252 & 1.601 & (d) & 3.285 & (b) & 2 & 3.263±0.097 \\ \hline
        EBLM J0520-06 & 2.13151 & - & - & 1.06 & (c)/(d) & 2.204 & (b) & 1 & 2.204±0.006 \\ \hline
        EBLM J0526+04 & 4.03101 & - & 4.5792 & - & - & - & - & 1 & 4.5792 \\ \hline
        EBLM J0546-18 & 3.19191 & - & - & 1.573 & (c)/(d) & 3.274 & (b) & 1 & 3.274±0.234 \\ \hline
        EBLM J0606-19 & 1.959995 & 1.96±0.038 & 1.9525 & 0.983 & (d) & 1.95 & (b) & 4 & 1.96±0.038 \\ \hline
        EBLM J0610-52 & 2.41699 & - & - & 1.211 & (c) & 2.433 & (b) & 10 & 2.433±0.182 \\ \hline
        EBLM J0621-46 & 1.55083 & 1.557±0.035 & 1.6003 & 0.778 & (d) & 1.583 & (b) & 6 & 1.557±0.035 \\ \hline
        EBLM J0621-50 & 4.96384 & 5.109±0.078 & 5.149 & 2.566 & (d) & 5.169 & (b) & 5 & 5.109±0.078 \\ \hline
        EBLM J0623-27 & 5.77793 & 5.819±0.045 & 5.8705 & 2.899 & (d) & 6.015 & (b) & 2 & 5.819±0.045 \\ \hline
        EBLM J0625-43 & 3.96899 & 3.984±0.052 & 4.0099 & 2.003 & (d) & 4.051 & (b) & 4 & 3.984±0.052 \\ \hline
        EBLM J0625-51 & 2.2061 & - & - & 1.104 & (c) & 2.21 & (b) & 6 & 2.21±0.118 \\ \hline
        EBLM J0640-27 & 2.92158 & - & - & 1.474 & (c) & 3.077 & (b) & 2 & 3.077±0.078 \\ \hline
        EBLM J0642-60 & 5.01154 & - & - & 2.499 & (c) & - & - & 19 & - \\ \hline
        EBLM J0814-20 & 3.13139 & - & - & 1.568 & (c)/(d) & - & (a) & 1 & - \\ \hline
        EBLM J0816+09 & 0.901455 & - & - & 0.451 & (c)/(d) & - & (a) & 1 & - \\ \hline
        EBLM J0855+04 & 2.22696 & - & - & 1.116 & (c) & - & (a) & 2 & - \\ \hline
        EBLM J0916-35 & 2.58835 & - & - & 1.297 & (c)/(d) & 2.457 & (b) & 1 & 2.457±0.0198 \\ \hline
        EBLM J0941-31 & 5.54563 & 5.375±0.04 & 5.3756 & 2.497 & (d) & 5.409 & (b) & 3 & 5.375±0.04 \\ \hline
        EBLM J0948-08 & 5.3798 & - & 5.3634 & 2.761 & (d) & 6.234 & (b) & 2 & 5.3633 \\ \hline
        EBLM J1007-40 & 3.93604 & - & - & 1.96 & (c) & - & (a) & 2 & - \\ \hline
        EBLM J1013+01 & 2.89228 & 3.299±0.092 & 3.3353 & 1.702 & NA & 3.28 & (b) & 4 & 3.299±0.092 \\ \hline
        EBLM J1016-42 & 4.3786 & 5.489±0.001 & - & 2.343 & NA & 5.237 & (b) & 1 & 5.489±0.001 \\ \hline
        EBLM J1023-43 & 3.68407 & - & - & 1.837 & (c) & - & (a) & 2 & - \\ \hline
        EBLM J1034-29 & 2.17426 & - & - & 1.099 & (c)/(d) & 2.178 & (b) & 1 & 2.178±0.017 \\ \hline
        EBLM J1037-25 & 4.93656 & 4.936±0.065 & 4.7236 & 2.539 & (d) & 4.97 & (b) & 1 & 4.936±0.065 \\ \hline
        EBLM J1037-45 & 1.59391 & - & - & 0.798 & (c) & 1.586 & (b) & 3 & 1.586±0.029 \\ \hline
        EBLM J1055-39 & 1.3516 & - & - & 0.678 & (c) & 1.373 & (b) & 3 & 1.373±0.045 \\ \hline
        EBLM J1104-43 & 1.76158 & - & - & 0.882 & (c) & - & (a) & 4 & - \\ \hline
        EBLM J1116-01 & 4.742 & 4.74±0.133 & 4.6974 & 2.328 & (d) & 4.707 & (b) & 3 & 4.74±0.133 \\ \hline
        EBLM J1141-37 & 5.14769 & 5.17±0.129 & 5.2749 & 2.655 & (d) & 5.14 & (b) & 3 & 5.17±0.129 \\ \hline
        EBLM J1208-29 & 2.67598 & 2.676±0.039 & 2.6584 & 1.334 & (d) & 2.712 & (b) & 2 & 2.676±0.039 \\ \hline
    \end{tabular}
\end{table*}

\begin{table*}
\caption{Table 1 continued.}
\label{rotation period catalog}
    \begin{tabular}{|c|c|c|c|cc|cc|c| c|}
    \hline
        EBLM Name & $P_{\rm orb}$ & $P_{\rm ACF}$ & $P_{\rm LS}$ & $1/2 P_{\rm bin}$ & Source & $P_{\rm bin}$ & Source & No. of TESS & Nominal $P_{\rm rot}$ \\ 
        & (days)&(days)&(days)&(days)& &(days)& & Sectors & (days) \\\hline
EBLM J1227-43 & 1.79554 & 1.799±0.031 & 1.8099 & 0.899 & (c)/(d) & 1.815 & (b) & 1 & 1.799±0.031 \\ \hline
        EBLM J1233-20 & 4.44384 & 4.694±0.067 & 4.7225 & 2.307 & (d) & 4.807 & (b) & 2 & 4.694±0.067 \\ \hline
        EBLM J1243-50 & 1.54691 & - & - & 0.765 & (d) & 1.611 & (b) & 1 & 1.612±0.014 \\ \hline
        EBLM J1350-31 & 3.12433 & - & - & 1.571 & (c)/(d) & 3.158 & (b) & 1 & 3.158±0.026 \\ \hline
        EBLM J1418-29 & 8.69122 & 7.841±0.124 & 8.8001 & - & (a) & 8.05 & (b) & 1 & 7.841±0.124 \\ \hline
        EBLM J1433-43 & 3.08248 & - & - & 1.547 & (c)/(d) & - & (a) & 1 & - \\ \hline
        EBLM J1543-27 & 2.57069 & - & - & 1.285 & (c)/(d) & - & (a) & 1 & - \\ \hline
        EBLM J1552-26 & 1.9492 & - & - & 0.977 & (c)/(d) & 1.986 & (b) & 1 & 1.986±0.021 \\ \hline
        EBLM J1907-45 & 5.6402 & 5.662±0.02 & 5.706 & 2.887 & (d) & 5.613 & (b) & 2 & 5.662±0.02 \\ \hline
        EBLM J1910-35 & 4.08311 & 4.081±0.044 & 4.0724 & 2.015 & (d) & 4.047 & (b) & 1 & 4.081±0.044 \\ \hline
        EBLM J2001-36 & 2.75353 & 2.75±0.128 & 2.6789 & 1.357 & (d) & 2.683 & (b) & 1 & 2.75±0.128 \\ \hline
        EBLM J2011-71 & 5.8727 & - & - & 2.97 & (c) & 5.895 & (b) & 2 & 5.895±0.18 \\ \hline
        EBLM J2017+02 & 1.49548 & - & - & 0.747 & (c) & 1.508 & (b) & 2 & 1.508±0.067 \\ \hline
        EBLM J2025-45 & 6.19199 & 6.429±0.079 & 6.3975 & 3.211 & (d) & 6.619 & (b) & 1 & 6.429±0.079 \\ \hline
        EBLM J2107-39 & 3.9618 & 3.986±0.063 & 4.0998 & 2.001 & (d) & 4.147 & (b) & 2 & 3.986±0.063 \\ \hline
        EBLM J2126-28 & 5.82329 & 8.385±0.129 & 8.2858 & - & (a) & 8.22 & (b) & 2 & 8.385±0.129 \\ \hline
        EBLM J2134-72 & 3.21407 & - & - & 1.615 & (c)/(d) & - & (a) & 1 & - \\ \hline
        EBLM J2153-55 & 8.54481 & 10.145±0.113 & 10.2471 & 4.959 & NA & 10.401 & (b) & 3 & 10.145±0.113 \\ \hline
        EBLM J2158-21 & 4.78197 & 4.818±0.028 & 5.5121 & 2.321 & (d) & 5.361 & (b) & 1 & 4.818±0.028 \\ \hline
        EBLM J2210-48 & 2.8201 & - & - & 1.405 & (c) & 2.892 & (b) & 2 & 2.892±0.156 \\ \hline
        EBLM J2232-31 & 3.14153 & 3.14±0.045 & 3.1831 & 1.583 & (d) & 3.195 & (b) & 2 & 3.14±0.045 \\ \hline
        EBLM J2236-36 & 3.06717 & 3.05±0.064 & 2.9508 & 1.53 & (d) & 3.064 & (b) & 2 & 3.05±0.064 \\ \hline
        EBLM J2308-46 & 2.19922 & 2.21±0.065 & 2.2239 & 1.107 & (d) & 2.195 & (b) & 2 & 2.21±0.065 \\ \hline
        EBLM J2309-67 & 1.95519 & 2.005±0.005 & 2.016 & 0.981 & (d) & 2.014 & (b) & 3 & 2.005±0.005 \\ \hline
        EBLM J2349-32 & 3.54967 & 4.232±0.048 & - & 2.083 & NA & 4.265 & (b) & 1 & 4.232±0.048 \\ \hline
        EBLM J0518-39 & 3.6498 & - & - & 1.793 & (c) & - & (a) & 3 & - \\ \hline
        EBLM J0500-35 & 8.28485 & 8.174±0.091 & 8.1999 & 4.156 & (d) & 8.239 & (b) & 5 & 8.174±0.091 \\ \hline
        EBLM J0649-27 & 4.30805 & - & - & 2.144 & (c) & 4.449 & (b) & 3 & 4.449±0.056 \\ \hline
        EBLM J0659-61 & 4.23505 & 4.308±0.066 & 4.3967 & 2.16 & (d) & 4.413 & (b) & 14 & 4.308±0.066 \\ \hline

    \end{tabular}
\end{table*}
\bsp	
\label{lastpage}
\end{document}